\documentstyle[12pt]{article}
\newcommand{\bce}{\begin{center}}
\newcommand{\ece}{\end{center}}
\newcommand{\beq}{\begin{equation}}
\newcommand{\eeq}{\end{equation}}
\newcommand{\bea}{\vspace{0.25cm}\begin{eqnarray}}
\newcommand{\eea}{\end{eqnarray}}

\newcommand{\ba}{\begin{array}}
\newcommand{\ea}{\end{array}}

\newcommand{\doublespace}{
    \renewcommand{\baselinestretch}{1.6}\large\normalsize}

\def\lsim{\mathrel{\rlap{\lower4pt\hbox{\hskip1pt$\sim$}}
    \raise1pt\hbox{$<$}}}	  %less than or approx. symbol
\def\gsim{\mathrel{\rlap{\lower4pt\hbox{\hskip1pt$\sim$}}
    \raise1pt\hbox{$>$}}}	  %greater than or approx. symbol

\def\lsim{\mathrel{\rlap{\lower4pt\hbox{\hskip1pt$\sim$}}
    \raise1pt\hbox{$<$}}}         %less than or approx. symbol
\def\gsim{\mathrel{\rlap{\lower4pt\hbox{\hskip1pt$\sim$}}
    \raise1pt\hbox{$>$}}}         %greater than or approx. symbol

  \advance\oddsidemargin  by -1in
  \advance\evensidemargin by -1in
  \advance\topmargin      by -0.5in
\def\lsim{\mathrel{\rlap{\lower4pt\hbox{\hskip1pt$\sim$}}
    \raise1pt\hbox{$<$}}}         %less than or approx. symbol
\def\gsim{\mathrel{\rlap{\lower4pt\hbox{\hskip1pt$\sim$}}
    \raise1pt\hbox{$>$}}}         %greater than or approx. symbol

\def\beq{\begin{equation}}
\def\endeq{\end{equation}}
\def\arr{\begin{eqnarray}}
\def\endarr{\end{eqnarray}}
\makeindex

\begin{document}

%\large

%{ \huge The last update:  24 January 1993\vspace{2.0cm}\\}
\phantom{.}{\bf \Large \hspace{9.4cm} KFA-IKP(Th)-1994-18\\
\phantom{.}\hspace{11.9cm}20 May 1994\\ }
%}
\begin{center}
{\huge \bf  Anomalous A-dependence of diffractive
leptoproduction of
radial   excitation
 $\rho'(2S)$\vspace{1.5cm}\\ }
{\Large \bf J.Nemchik$^{1,3)}$,
N.N.Nikolaev$^{2,3}$
and B.G.Zakharov$^{2}$\bigskip\\ }
{\sl
$^{1)}$Institute of Experimental Physics, Slovak Academy of
Sciences, \\Watsonova 47, 043 53 Kosice, Slovak Republik   \\
$^{2)}$L.D.Landau Institute for Theoretical Physics, GSP-1, 117940,\\
ul.Kosygina 2, V-334, Moscow, Russia\\
$^{3)}$IKP(Theorie), KFA J\"ulich, D-52425 J\"lich, Germany
\vspace{0.4cm}\\ }

{\Large \bf A b s t r a c t}
\smallskip\\
\end{center}

We predict a strikingly different $A$ and $Q^{2}$ dependence of
quasielastic
leptoproduction of the $\rho^{0}(1S)$-meson and its radial
excitation $\rho '(2S)$ on nuclei.
Whereas for the $\rho^{0}$ production
nuclear transparency $T_{A}$
decreases monotonically with $A$, for the
$\rho'$ nuclear transparency $T_{A}$ can have the counterintuitive
nonmonotonic $A$-dependence, having the minimum for light nuclei,
and increasing with $A$ for medium and heavy nuclei.
Strong enhancement of the $\rho ' /\rho^{0}$ cross section ratio
makes nuclear targets the $\rho '$-factory.
The origin
of the anomalous $A$-dependence is in the interplay of color
transparency effects with the nodal structure of the $\rho '$
wave function. The predicted effects take place at moderate
$Q^{2}\lsim (2-3)$GeV$^{2}$, which can be
explored in high statistics experiments at CEBAF.
\vspace{0.5cm}\\
\centerline{\sl Submitted to Physics Letters B}
\centerline{E-mail: kph154@zam001.zam.kfa-juelich.de}
\pagebreak

\doublespace
%----------------------------------------------------------------------

%                              Section 1

%----------------------------------------------------------------------

%\section{Introduction}

The recent FNAL E665 experiment [1] on the exclusive
virtual photoproduction of the $\rho^{0}$-mesons
on nuclei [1] confirmed the  long anticipated effect of color
transparency (CT): decrease of nuclear attenuation of the produced
$\rho^{0}$ mesons with increasing virtuality $Q^{2}$ of photons
[2,3]. Regarding the accuracy of the data, the potential
of the CERN and FNAL muon scattering experiments is nearly
exhausted. Fortunately, very high statistics experiments can
be performed at CEBAF upgraded to the 8-12 GeV energy range [4].
(For discussion of the vector meson physics as the case for
10-20 GeV electron facility see [5], high energy aspects of
leptoproduction of vector mesons are discussed in [3,6], for
the recent review on CT see [7].)
Apart from the much more detailed studies of the onset of CT
in the $\rho^{0}$ production,
the CEBAF experiments can open an entirely new window on CT,
as due to high luminocity and excellent CLAS facility,
an accurate measurement of production of the
radially excited $\rho'$-meson becomes possible for the first
time. As we shall demonstrate below, CT leads
to a spectacular pattern of anomalous $A$ and $Q^{2}$ dependence
of the $\rho'$ production.

The crucial observation is that the very mechanism of
CT leads to a novel phenomenon of scanning the
wave function of vector mesons [2,8,9]. Specifically, the
amplitude of the free-nucleon reaction
$\gamma^{*}+N\rightarrow V+N$ can be written as
\beq
M_{N}= \int d^{2}\vec{r} \sigma(r)
\int_{0}^{1}dz
\Psi^{*}_{V}(z,r)
\Psi_{V}(z,r) \, .
\label{eq:1.1}
\endeq
Here we use the wave functions of the $q\bar{q}$ Fock states of
the vector meson and the photon in the lightcone representation
[10], where $\vec{r}$ is the transverse separation of the quarks, and
$z$ is a fraction of the lightcone momentum carried by
the quark. In Eq.~(\ref{eq:1.1}), $\sigma(r)$ is the cross section
for interaction with the nucleon of the $q\bar{q}$ color dipole
of size $r$. By virtue of CT, for the small-size
color dipole [10,11]
\beq
\sigma(r) \propto r^{2}\, ,
\label{eq:1.2}
\endeq
and the production amplitude (\ref{eq:1.1}) receives the dominant
contribution from ([3] and see below)
\beq
r \sim r_{S} = { 6 \over \sqrt{m_{V}^{2}+Q^{2}}} \, .
\label{eq:1.3}
\endeq

The wave function of the radial excitation
$\rho'(2S)$ has a node. For this reason, in the $\rho'$
production amplituide there is the node efffect -
 cancellations between the
contributions from $r$ below, and above, the node, which
depends on the scanning radius
$r_{S}$. If the node effect is
strong, even the slight variations of $r_{S}$
lead to an anomalously rapid variation of the $\rho'$ production
amplitude, which must be contrasted to the smooth $Q^{2}$ and
$r_{S}$ dependence of the $\rho^{0}(1S)$
production amplitude.
The point which we wish to make in this
communication is that apart from changing $Q^{2}$,
 one can also vary the effective scanning
radius employing a CT property of
stronger nuclear attenuation of the
large-$r$ states. We predict
a strikingly different, and often counterintuitive,
$A$ (and $Q^{2}$) dependence
of the $\rho '$ and $\rho^{0}$ production on nuclei, the
experimental observation of which will shed new light on our
understanding of the mechanism of CT.

%----------------------------------------------------------------------

%                                 Section 2

%----------------------------------------------------------------------

%\section{Scanning the wave function of the $\rho '$-meson}

We start with the derivation of the scanning radius $r_{S}$
Eq.~(\ref{eq:1.3}).
The most important feature of the photon wave function
$\Psi_{\gamma^{*}}(r,z)$ is an exponential decrease at large
distances [2,3,8-10]
%--------------------------------------
\begin{equation}
\Psi_{\gamma^{*}}(r,z)
\propto \exp(-\varepsilon r)   \, ,
\label{eq:2.1}
\end{equation}
%--------------------------------------
where
%--------------------------------------
\begin{equation}
\varepsilon^{2} = m_{q}^{2}+z(1-z)Q^{2} \, .
\label{eq:2.2}
\end{equation}
%--------------------------------------
In the nonrelativistic quarkonium $z\sim {1\over 2}$, and
the relevant $q\bar{q}$ fluctuations have a size
%-------------------------------------
\begin{equation}
r \sim r_{Q} = {1 \over \sqrt{m_{q}^{2}+{1\over 4}Q^{2}}}
\approx {2 \over \sqrt{m_{V}^{2}+Q^{2}}} \, .
\label{eq:2.3}
\end{equation}
%-------------------------------------
The wave function of the vector meson is smooth at small $r$.
Then, because of CT property Eq.~(\ref{eq:1.2}), the integrand
in Eq.~(\ref{eq:1.1})  will be peaked at $r\sim r_{S}\approx
3 r_{Q}$ as it is stated in Eq.~(\ref{eq:1.3}).
The product
$\sigma(r)\Psi_{\gamma^{*}}(z,r)$ acts as a distribution which
probes the wave function of the vector meson at the scanning
radius $r\sim r_{S}$ [2,9].
The large numerical factor
in the {\sl r.h.s.} of Eq.~(\ref{eq:1.3}) stretches
the large-size dominance, and
the simple nonrelativistic approximation
$z \sim {1\over 2}$ remains viable in a broad range of
$Q^{2} \lsim (3-5)$GeV$^{2}$ of the interest at CEBAF. (The
including of the
relativistic effects is straightforward, does not change the
principal results,
 and shall not
be discussed here.)
In Fig.1 we show qualitatively, how this
$Q^{2}$-dependent scanning works for the $\rho^{0}$ and $\rho '$
mesons. In our numerical calculations
we use the dipole cross section $\sigma(r)$ of Ref.~[10]
and the simple
harmonic oscillator model with the quark mass $m_{q}=0.3$GeV
and $2\hbar \omega =0.7$GeV. The radius of the $\rho^{0}$-meson
is essentially identical to that of the pion, and within the model
$\sigma_{tot}(\rho^{0} N) \approx \sigma_{tot}(\pi N) \approx 25$mb.

For the $\Psi'$ and
$\Upsilon'$ production, the node effect
is perturbatively tractable even for the
real photoproduction $Q^{2}=0$. The prediction
[8] of $\sigma(\gamma N\rightarrow \Psi' N)/
\sigma(\gamma N\rightarrow J/\Psi N)=0.17 $ is in excellent
 agreement
with the NMC result $0.20 \pm 0.05(stat) \pm 0.07(syst)$ for this
ratio [12]. In this case the node effect is relatively weak.
For the light mesons the scanning radius $r_{S}$ is larger
and, at small $Q^{2}$, the node effect
is much stronger. With the above described harmonic oscillator
wave function, we find the $Q^{2}$ dependence of the $\rho'/\rho^{0}$
production ratio shown in Fig.2. It corresponds to the
{\sl overcompensation} scenario when,
at $Q^{2}=0$, the $\rho'$ production
amplitude is dominated by the contribution from $r$ above the node.
As $Q^{2}$ increases and the scanning radius $r_{S}$ decreases, we
encounter the exact node effect, the $\rho'/\rho^{0}$ production ratio
has a dip at finite $Q^{2}$, and
with the further increase of $Q^{2}$ and decrease of
$r_{S}$, we have the {\sl undercompensation} regime -
the free-nucleon amplitude will be
dominated by the contribution from $r$ below the node.
The node effect will decrease with rising $Q^{2}$,
see the pattern of the scanning
shown in Fig.~1, and the $\rho'/\rho^{0}$ ratio will
rise with $Q^{2}$.

At small $Q^{2}$ the scanning radius is large, and we are not in
the domain of the perturbative QCD. What, then, are our firm
predictions? Firstly, the strong node effect and the strong
suppression of the real photoproduction of the $\rho'$ is not
negotiable, and this prediction is consistent with the meagre
experimental information (for the review see [13]). Secondly,
the wave functions of the $\rho^{0}(1S)$ and $\rho'(2S)$
at the origin are approximately equal. Therefore,
at very large $Q^{2}$ when $r_{S}$  is very small,
we have a firm prediction that
$d\sigma_{\rho '} \sim  d\sigma_{\rho^0}$. In the above
overcompensation scenario, the
$\rho'/\rho^{0}$ ratio has a dip at finite
$Q^{2}$. The position of this dip is
model dependent, but the prediction of the steep rise of the
$\rho'/\rho^{0}$ in the region of $Q^{2}\lsim (2-3)$GeV$^{2}$ is
a firm consequence of the $Q^{2}$-dependence of the scanning
radius, which is driven by CT. This is precisely the region
where the high statistics data can be taken at CEBAF, and
the experimental observation of such a dramatic
large-distance manifestation of CT will be very important
contribution to our understanding of the onset of CT.

%----------------------------------------------------------------------
%                               Section 3
%----------------------------------------------------------------------

%\section{The two possible scenarios for the node effect in the
%$\rho '$ production on nuclei}

The $\rho'$ production on nuclei is indispensable for testing
the node affect, as nuclear attenuation gives still another
handle on the scanning radius.
For the sake of definiteness, we discuss the quasielastic
(incoherent) $\rho'$ production on nuclei, extension to the
coherent production is straightforward and will be presented
elsewhere.

At high energy $\nu$, the
(virtual) photon forms its $q\bar{q}$ Fock state at a distance
(the coherence length) in front of the target nucleus (nucleon)
%-------------------------------------------------
\begin{equation}
l_{c}={{2\nu} \over Q^{2}+m_{V}^{2}} \, .
\label{eq:3.1}
\end{equation}
%-------------------------------------------------
After interaction with the target, the $q\bar{q}$ pair
recombines into the observed
vector meson $V$ with the formation (recombination)
length
%-------------------------------------------------
\begin{equation}
l_{f}={\nu \over m_{V}\Delta m} \, ,
\label{eq:3.2}
\end{equation}
%-------------------------------------------------
where $\Delta m $ is the typical level splitting in the
quarkonium. At low energy and $l_{f}\ll R_{A}$,
where $R_{A}$ is a radius of the target nucleus,
recombination
of the $q\bar{q}$ pair into the vector meson takes place well
inside the nucleus, nuclear attenuation will be given by the
free-nucleon $VN$ cross section and CT effect disappears.
The condition $l_{f}\gsim R_{A}$ is
crucial for the onset of CT,
and for the $\rho^{0},\rho '$ system it requires
\beq
\nu \gsim (3-4)\cdot A^{1/3}\,{\rm GeV} \, ,
\label{eq:3.3}
\endeq
which for light and medium nuclei
is well within the reach of the $8-12$GeV upgrade of CEBAF.
(In this energy range we can also limit ourselves to the
contribution from the lowest $q\bar{q}$ Fock states of the
photon and vector meson.)
In the opposite to that, the condition $l_{c} \gsim R_{A}$ is not
imperative for the onset of CT. Nuclear attenuation effects
only increase somewhat when $l_{c}$ increases from $l_{c}\ll R_{A}$
to $l_{c}\gsim R_{A}$ [2,14]. The
finite-energy effects can easily be incorporated
using the path integral technique developed in [8,2].
For the sake of simplicity, in this
paper we concentrate on the high-energy limit of
$f_{f},l_{c} \gsim R_{A}$, when
nuclear transparency
$T_{A} = d\sigma_{A}/A\sigma_{N}$ equals [2,9,14]
\arr
Tr_{A}={1\over A}
\int d^{2}\vec{b} T(b)
{\langle V |\sigma(r)
\exp\left[-{1\over 2} \sigma(r)T(b)\right] |\gamma^{*}
\rangle^{2} \over
\langle V|\sigma(r)|\gamma^{*}\rangle^{2} }    \, .
\label{eq:3.4}
\endarr
%--------------------------------------
Here $T(b) = \int dz n_{A}(b,z)$ is the optical thickness of
a nucleus, where $n_{A}(b,z)$ is the nuclear matter density
(For the compilation of the nuclear density parameters see [15]).
The $A$-dependence of the node effect comes from the
nuclear attenuation factor
$\exp[-{1\over 2} \sigma(r)T(b)]$ in the nuclear matrix
element $M_{A}(T)=
 \langle V |\sigma(r)
\exp\left[-{1\over 2} \sigma(r)T(b)\right] |\gamma^{*}
\rangle$.

Firstly, consider
the value of $Q^{2}$, at which the cross section
for the $\rho'$ production on the free nucleon
takes its minimal value because of the exact
node effect. Because of the $r$-dependence of the attenuation
factor, in the nuclear amplitude the node effect will
be incomplete. Consequently, as a function of $Q^{2}$, nuclear
transparency $T_{A}$ will have a spike $T_{A} \gg 1$ at a
finite value of $Q^{2}$ [2].

Secondly, consider the $\rho '$ production on nuclei at a
fixed value of $Q^{2}$ such that the
free nucleon amplitude is still in the overcompensation regime.
Increasing $A$ and enhancing the importance of the attenuation
factor $\exp[-{1\over 2} \sigma(r)T(b)]$,
we shall bring the nuclear amplitude
to the nearly exact compensation regime. Therefore, the
$\rho'/\rho^{0}$ production ratio, as well as nuclear transparency
for the $\rho'$ production,  will decrease with $A$ and
take a minimum value at a certain finite
$A$. With the further increase of $A$, the undercompesation regime
takes over, and we encounter very counterintuitive situation:
nuclear transparency for the $\rho'$ is larger for the strongly
absorbing nuclei!  This situation is illustrated in Fig.~3a and
must be contrasted with a
smooth and uneventful decrease of transparency
for the $\rho^{0}$ production on heavy nuclei.

Evidently, the possibility of the perfect node effect in $M_{A}(T)$
depends on the optical thickness $T(b)$.
This is shown in Fig.~4, in which we present the
relative nuclear matrix element
%--------------------------------------
\beq
R(T)={ \langle \rho'|\sigma(r)
\exp[-{1\over 2}\sigma(r)T(b)]|
\gamma^{*}\rangle
\over \langle \rho'|\sigma(r)|\gamma^{*}\rangle } \, .
\label{eq:3.5}
\endeq
%--------------------------------------
At large impact parameter $b$, at the periphery of the nucleus, the
optical thickness of the nucleus is small, the
overcompensation in the nuclear matrix element
is the same as for the free nucleon, and we
have $R(T)=1$. At smaller impact parameters, one encounters the
exact node effect: $R(T) =0$. At still smaller impact parameters,
the overcompensation changes for the undercompensation and
$R(T) < 0$. Here the breaking of the compensation and the
overall attenuation start competing. For very heavy nuclei and
small impact parameters, the overall attenuation takes over
and $R(T)$ starts decreasing again. For the particular case
of the real photoproduction, and for the specific wave function
of the $\rho '$ meson, in the undercompensation domain
we have $|R(T)| \leq 1$. The resulting $A$ dependence of nuclear
transparency $T_{A}$ is shown in Fig.~3a. It takes the minimum
value for the $^{7}Li$ target, then increases with $A$, flattens
and starts decreasing for very heavy nuclei.

In Fig.~3 we show how nuclear transparency for the $\rho'$ varies
with $Q^{2}$.
The slight increase of $Q^{2}$ up to $Q^{2}=0.2$GeV$^{2}$ and
the slight change of the
scanning radius enhance the node effect in the free nucleon
amplitude, see Fig.~2.
 In this case, we find almost exact node effect
for the $^{4}He$ target, which is followed by the dominant
undercompensation regime for heavier nuclei (Fig.~3b). Also,
in this case
the undercompensation regime for heavy nuclei is followed by
$|R(T)| > 1$, which leads to significant
antishadowing $T_{A} > 1$.

With the further increase of $Q^{2}$ one enters the pure
undercompensation regime for all the targets.
Nuclear suppression of the node effect enhances $M_{A}$
and nuclear transparency $T_{A}$, whereas the overall
attenuation factor
$\exp[-{1\over 2}\sigma(r)T(b)]$ decreases $T_{A}$. Of these
two competing effects, the former remains stronger and we
find antishadowing of the $\rho'$ production in a broad
range of $A$ and $Q^{2}$, see Figs.~3c-e.
Typically, we find a nuclear enhancement
of the $\rho'/\rho^{0}$ production ratio on heavy targets
by one order in the
magnitude with respect to the free nucleon target. This makes
leptoproduction on nuclei the $\rho'$ factory, and the $\rho'$
production experiments at CEBAF can contribute much to the
poorly understood spectroscopy of the radially excited vector
mesons.
Only at a relatively large
$Q^{2}\gsim 2$GeV$^{2}$, the attenuation effect takes over,
and
nuclear transparency for the $\rho '$ production will
start decreasing monotonically with $A$ (Fig.~3f).
Still, this decrease
is much weaker
than for the $\rho^{0}$ meson.
At very large $Q^{2}$, when the node effect disappears
because of the small scanning radius $r_{S}$, nuclear transparency
for the $\rho^{0}$ and the $\rho '$ production will become
identical. This pattern repeats qualitatively the one studied
in [2,8] for the $J/\Psi$ and $\Psi'$ mesons.

Notice, that for the $\rho^{0}$-production, the nuclear
transparency is a very slow function of $Q^{2}$. For instance,
for the lead target $T_{Pb}(Q^{2}=0)\approx 0.1$ and
$T_{Pb}(Q^{2}= 2{\rm GeV}^{2})\approx 0.23$. The reason for
this slow onset of color trasnparency is that because of
large numerical factor in the r.h.s of Eq.~(\ref{eq:1.3}),
even at $Q^{2}=2{\rm GeV}^{2}$ the scanning radius is still
large, $r_{S}\sim  0.8$f. Furthermore, the
more detailed analysis [3] has shown
that nuclear shadowing is controlled by a still larger
$r \sim 5r_{Q}$. Our numerical predictions [2,3] for the $Q^{2}$
dependence of the incoherent and coherent $\rho^{0}$ production
on nuclei are in excellent agreement with the E665 data.

The above presented results refer to the production of the
transversely polarized $\rho^{0}$ and $\rho'$ mesons. Accurate
separation of production of the transverse and longitudinal
cross section can esily be done in the high statistics CEBAF
experiments. Here we
only wish to mention the interesting possibility that for the
longitudinally polarized $\rho'$ mesons, the exact node effect
can take place at a value of $Q^{2}$ different from that for the
transversely polarized $\rho '$ mesons, and polarization of
the produced $\rho'$ can exhibit very rapid change with $Q^{2}$.
\smallskip\\
%----------------------------------------------------------------------
%                               Conclusions
%----------------------------------------------------------------------

{\large \bf Discussion of the results and conclusions:}\smallskip\\
We presented the strong case for the anomalous $Q^{2}$ and
$A$ dependence of incoherent production of the $\rho '$
meson on nuclear targets. The origin of the effect is in
the $Q^{2}$ dependence of the scanning radius $r_{S}$,
which follows from color transparency property in QCD.
At the relatively small values of $Q^{2}$ discussed in this
paper, the scanning radius is rather large, of the order of
the size of the $\rho^{0}$ meson. For this reason, the
$\rho '$ production amplitude proves to be extremely sensitive
to the nodal structure of the $\rho '$ wave function.
Specifically, the node effect leads to a strong suppression
of the $\rho' /\rho^{0}$ ratio in the real photoproduction,
which is consistent with the meagre experimental information
on the production of radially excited vector mesons.

In the overcompensation scenario suggested by the
nonrelativistic oscillator model, the most striking effect is the
nonmonotonic $A$-dependence of nuclear transparency shown in
Fig.~3. The numnerical
predictions are very sensitive to the position of the node in
the wave function of $2S$ states.
It is quite possible that the dip of nuclear transparency $T_{A}$
will take place for targets
much heavier than in Figs.~3a,b, and disappearance of the node
effect and the onset of the more conventional
nuclear shadowing $T_{A} <1$ for the $\rho '$ production
like in Fig.~3f only will take place at much larger $Q^{2}$.
Also, the possibility of the undercompensation at $Q^{2}=0$
can not be excluded.
However, the strikingly different $A$-dependence of the
incoherent $\rho^{0}$ and $\rho'$ production on nuclei persists
in such a broad range of $Q^{2}$ and of the scanning radius
$r_{S}$, that the existence of the phenomenon of anomalous
$A$ and $Q^{2}$ dependence of the $\rho'$ preduction is not
negotiable. It is a direct manifestation of
the color-transparency driven $Q^{2}$ dependence of the
scanning radius and, as such, it deserves a dedicated experimental
study.

For the sake of simplicity, here we concentrated on the high
energy limit in which the  theoretical treatment of nuclear
production greatly simplifies. One can easily go beyond the
frozen-size approximation employing  the path-integral technique
[2,8]. The detailed analysis performed in  [2] shows
that at energies of the virtual photon $\sim 6-10$GeV, the
subasymptotic corrections do not change much the predictions for
nuclear transparency. This is the energy range wich can be
reached at CEBAF after the energy upgrade.

Few more comments about the possibilities of CEBAF are worth
while. Because of the strong suppression of the $\rho'/\rho^{0}$
production ratio, the high luminosity of CEBAF is absolutely
crucial for high-statistics experiments on the $\rho'$ production.
Notice, that the most interesting anomalies in the $A$ and
$Q^{2}$ dependence take place near the minimum of the $\rho'$
production cross section. Furthermore, the observation of the
$\rho'$ production requires detection of its 4-pion decays,
and here one can take advantage of the CLAS multiparticle
spectrometer available at CEBAF.
Finally, similar effects must persist also for the $\omega'$ and
$\phi'$ production, and also in the coherent production of the
radially excited mesons.\smallskip\\
{\large \bf Acknowledgements:} \smallskip\\
One of the authors (J.N.) is grateful to J.Speth for the
hospitality at Institut f. Kernphysik, KFA J\"ulich, where this
work was completed. N.N.N. thanks R.Holt and
N.Isgur for discussions during
the workshop on CEBAF at Higher Energies.

\pagebreak

{\large \bf Figure captions:}

\begin{itemize}
\item
Fig.1. -
The qualitative pattern of the the $Q^{2}$-dependent scanning
of the wave functions of the ground state $V(1S)$ and
the radial excitation
$V'(2S)$ of the vector meson. The scanning distributions
$\sigma(r)\Psi_{\gamma^{*}}(r)$ shown
by the solid and dashed
curve have the scanning radii $r_{S}$ differing by a factor 3.
The wave function and radius $r$ are in arbitrary units.

\item
Fig.2. -
The $Q^{2}$ dependence of the $\rho'(2S)/\rho^{0}(1S)$ ratio of
forward
production cross sections.

\item
Fig.3. -
The $Q^{2}$ and $A$ dependence of nuclear transparency for the
$\rho^{0}(1S)$ and $\rho'(2S)$ electroproduction on nuclei.

\item
Fig.4. -
The impact parameter dependence of the reduced nuclear
matrix element $R(T)$ for the real photoproduction of the
$\rho'(2S)$. \\

\end{itemize}

\pagebreak

\end{document}